\documentclass[aps,prl,twocolumn,superscriptaddress]{revtex4}

\usepackage{color}
\usepackage{graphicx}

\begin{document}

\input epsf.sty
\title{Pressure-induced commensurate 
magnetic order in multiferroic HoMn$_{2}$O$_{5}$}


\author{H. Kimura}
\email[]{kimura@tagen.tohoku.ac.jp}

\author{K. Nishihata}

\author{Y. Noda}
\affiliation{Institute of Multidisciplinary Research for Advanced Materials, 
Tohoku University, Sendai 980-8577, Japan}

\author{N. Aso}
\affiliation{Institute for Solid State Physics, The University of Tokyo, Tokai, Ibaraki 319-1106, Japan}

\author{K. Matsubayashi}

\author{Y. Uwatoko}
\affiliation{Institute for Solid State Physics, The University of Tokyo, Kashiwa 277-8581, Japan}

\author{T. Fujiwara}
\affiliation{Graduate School of Science and Engineering, Yamaguchi University, Yamaguchi 753-8512, Japan}


\date{\today}

\begin{abstract}
The pressure ($p$) -- temperature ($T$) phase diagram for microscopic magnetism 
in the multiferroic compound HoMn$_{2}$O$_{5}$ was established 
using neutron diffraction measurements under a hydrostatic pressure 
up to 1.25~GPa. At ambient pressure,
incommensurate--commensurate--incommensurate magnetic phase transitions 
occur successively with decreasing temperature.
Upon applying pressure, the incommensurate 
phase at the lowest temperature 
almost decreases and the commensurate phase appears. 
The $p$ -- $T$ phase diagram established shows excellent agreement with 
the recently reported $p$ -- $T$ dielectric phase diagram, where 
ferroelectricity is induced by applying pressure. 
We also found that the $p$ -- $T$ magnetic phase diagram is quite similar 
to the previously obtained magnetic field-temperature phase diagram. 
\end{abstract}

\pacs{61.12.Ld, 64.70.Rh, 75.47.Lx, 75.80.+q}

\maketitle

\begin{figure}[hbp]
\begin{center}
\includegraphics[width=80mm]{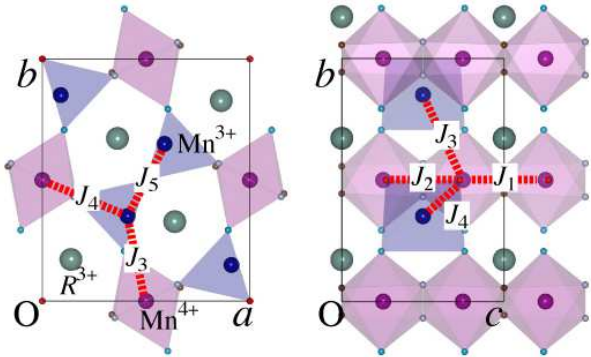}
\end{center}
\caption{
(Color online) 
Crystal structure of $R$Mn$_{2}$O$_{5}$ 
at paraelectric phase projected along the $ab$-plane (left panel) and 
$bc$-plane (right panel). $J_{1}\sim J_{5}$ indicates possible exchange interactions
acting on 
neighboring manganese spins. 
}
\label{fig1}
\end{figure}

The coexistence of and 
spontaneous ordering of 
(anti) ferromagnetism and 
ferroelectricity, 
referred to as multiferroics, has recently been identified in certain materials. 
Both the scientific and technological aspects of these materials have attracted much attention owing to the possible 
colossal magnetoelectric (ME) effect, in which the 
electric polarization can be controlled by a magnetic field, or 
conversely, the magnetization can be controlled by an electric field. 
The series of rare-earth manganese compounds $R$Mn$_{2}$O$_{5}$ ($R=$ rare earth, Bi, and Y) 
is a prototype multiferroic system that exhibits the 
colossal ME effect\cite{Hur2004-1,Hur2004-2}. As shown in Fig.~\ref{fig1}, 
this system has two independent Mn sites of Mn$^{3+}$ and Mn$^{4+}$ ions, a network of which 
surrounds the $R^{3+}$ ion. Furthermore, there are at least five different paths 
($J_{1}\sim J_{5}$ in Fig.~\ref{fig1}) of Mn-Mn 
exchange interactions\cite{Blake2005}. These geometrical configurations potentially involve 
magnetic frustration, resulting in a complex phase sequence of multiple 
magnetic transitions as a function of temperature\cite{Kimura2007-1}. 
Magnetic frustration has recently received interest in connection with
multiferroics because many multiferroic 
materials, such as $R$MnO$_{3}$\cite{Kimura2003}, 
Ni$_{3}$V$_{2}$O$_{8}$\cite{Lawes2005}, and MnWO$_{4}$\cite{Taniguchi2006}, 
essentially involve geometrical magnetic frustration.
The competition between multiple magnetic ground states within a small energy scale 
due to the frustration allows various magnetic phase transitions to occur upon the application of an external field. 
A rich variety of magnetic as well as dielectric responses to 
the applied magnetic field has been found in 
$R$Mn$_{2}$O$_{5}$ with $R=$ Tb\cite{Hur2004-1,Hur2004-2}, 
$R=$ Dy\cite{Hur2004-2,Higashiyama2004,Ratcliff2005}, 
$R=$ Ho\cite{Hur2004-2,Higashiyama2005,Kimura2006}, 
and $R=$ Er\cite{Higashiyama2005,Kimura2007-2}.
Although the origin of these magnetic field responses have been discussed
on the basis of the interaction between the applied magnetic field and 
the induced rare-earth magnetic moment\cite{Kimura2007-3},
the details are not yet understood.

It was recently found that a hydrostatic pressure of around 1~GPa can
have a dramatic effect on the dielectric properties
of Ni$_{3}$V$_{2}$O$_{8}$ and $R$Mn$_{2}$O$_{5}$: ferroelectricity is suppressed in the former, \cite{Chaudhury2007} 
while in the latter system,
ferroelectricity is restored\cite{Cruz2007} by applying pressure. 
In contrast to the case of an external magnetic field,
where the magnetic field can directly couple to spins, pressure
can tune the magnetic interaction between neighboring
spins by decreasing the interatomic distance, and by changing the bond angles. 
A small change in the arrangement around magnetic ions causes 
magnetic phase transitions because of the magnetic frustration, 
which yields dielectric phase transitions in multiferroic systems. 
However, no microscopic evidence of pressure-induced magnetic phase 
transitions has yet been found in any multiferroic system. 
Thus, in the present study, we performed neutron diffraction measurements
under hydrostatic pressure in the multiferroic HoMn$_{2}$O$_{5}$ system in
search for microscopic magnetic responses to pressure and to determine 
the relationship between the magnetic response and the dielectric response. 
\begin{widetext}
\begin{figure}[hbtp]
\begin{center}
\includegraphics[width=165mm]{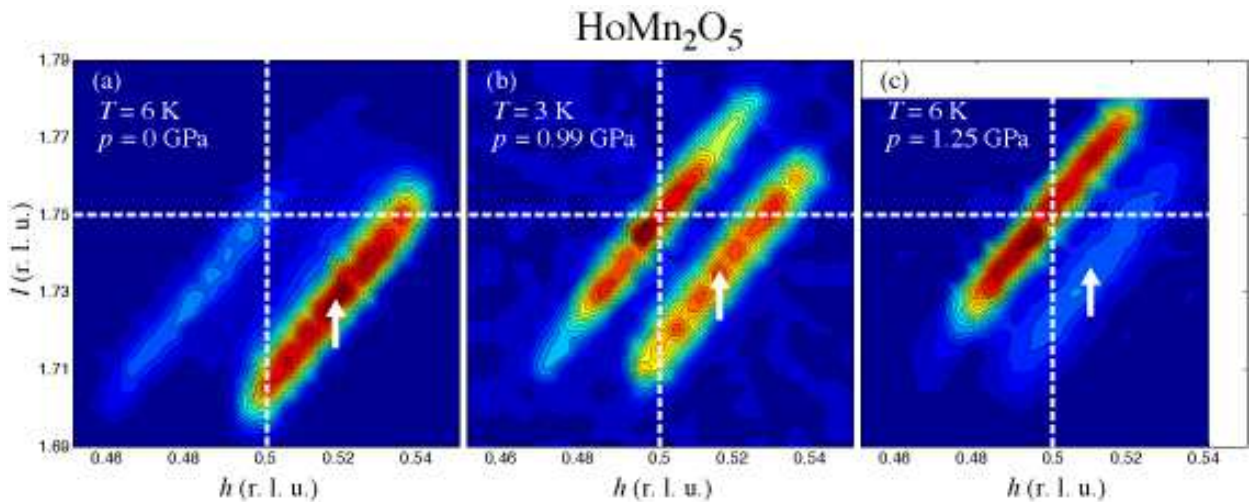}
\end{center}
\caption{
(Color online) 
Contour maps of the magnetic Bragg reflections for HoMn$_{2}$O$_{5}$ 
around (0.5\ 0\ 1.75) in the 
($h\ 0\ l$) zone below $T=6$~K upon applying a
hydrostatic pressure of (a) 0~GPa, (b) 0.99~GPa, and (c) 1.25~GPa. 
The white dashed lines denote the commensurate position at $h=1/2$ and $l=7/4$. 
The white arrows indicate the center position of incommensurate signals.
}
\label{fig2}
\end{figure}
\end{widetext}
We successfully established the pressure ($p$) -- temperature ($T$) phase diagram for 
microscopic magnetism and found a strong correlation to
the $p$ -- $T$ dielectric phase diagram reported previously\cite{Cruz2007}. 

A single crystal of HoMn$_{2}$O$_{5}$ was grown by the
PbO--PbF$_{2}$ flux method\cite{Wanklyn1972}. 
The crystal, which was almost a cube, $1.4$~mm on a side, was inserted
in a pressure cell. A hydrostatic pressure of up to 1.25~GPa was applied using a
copper-beryllium (CuBe) based piston-cylinder clamp device\cite{Aso2006} 
with a (1:1) mixture of Fluorinert FC75/77 as a pressure transmitting medium. 
The applied pressure was measured by evaluating the pressure change in the
superconducting transition temperature of high-purity lead, 
which was placed in the pressure cell with the HoMn$_{2}$O$_{5}$ crystal. 
Neutron diffraction measurements were performed using the
thermal neutron triple-axis spectrometer 
AKANE owned by the Institute of Material Research of Tohoku University, 
installed at JRR-3 in the Japan Atomic Energy Agency. 
The sample was mounted on the ($h\ 0\ l$) 
scattering plane. The incident energy of neutrons was fixed at 
19.5~meV using a Germanium (311) monochromator.

HoMn$_{2}$O$_{5}$ is orthorhombic with $Pbam$ symmetry for the paraelectric phase. 
As shown in Fig.\ref{fig1}, edge-sharing Mn$^{4+}$O$_{6}$ octahedra align along 
the $c$-axis and pairs of Mn$^{3+}$O$_{5}$ pyramids link the Mn$^{4+}$O$_{6}$ chains 
in the $ab$-plane, a network of which surrounds the Ho$^{3+}$ ion\cite{Abrahams1967}. 
At ambient pressure, HoMn$_{2}$O$_{5}$ show
successive magnetic phase transitions: below a N$\acute{\rm e}$el temperature $T_{\rm N1}\sim 44$~K, 
a high-temperature incommensurate magnetic (HT-ICM) phase appears 
in the absence of electric polarization. 
At  $T_{\rm CM}\sim 39$~K, the HT-ICM phase disappears and 
the commensurate magnetic (CM) phase arises, 
where an electric polarization is concurrently induced along the $b$-axis. 
As the temperature decreases, a further transition occurs from the CM phase to a
low-temperature incommensurate magnetic (LT-ICM) phase at $T_{\rm N2}\sim 20$~K,
at which temperature a dielectric phase transition occurs as well, the electric polarization
drops, and the system is converted into an unidentified dielectric phase (X phase). 
All the magnetic phases are characterized by a magnetic 
propagation wave vector ${\bf q}_{\rm M}=(q_{x}\ 0\ q_{z})$, where the periodicity 
of the magnetic order along both the $a$- and $c$-axes is modulated. 
Details of the magnetic and dielectric properties under ambient pressure
have been reported elsewhere\cite{Kimura2006}. 

Under hydrostatic pressure, ${\bf q}_{\rm M}$ observed by neutron diffraction 
significantly changes. Figure~\ref{fig2} displays contour maps of the magnetic Bragg reflections 
around (0.5\ 0\ 1.75) sliced in the $(h\ 0\ l)$ reciprocal zone as a function of applied pressure,
taken below 6~K. All the peak profiles are elongated in the diagonal direction owing to
the instrument resolution. Under ambient pressure (see Fig.~\ref{fig1}(a)), the pair of 
incommensurate peaks observed around (0.5$\pm$0.02\ 0\ 1.73) are well-defined, which indicates
the LT-ICM phase. The imbalance in the intensity 
of 
the two peaks results from
the domain distribution or magnetic structure factor. As the applied pressure is increased to 
$p=0.99$~GPa, the incommensurate peak becomes weak and a commensurate peak appears at 
$Q=(1/2\ 0\ 7/4)$ (denoted by white dashed lines in the figure), where the LT-ICM and CM phases coexist. At $p=1.25$~GPa (see Fig.~\ref{fig1}(c)),
the commensurate peak is dominant, with a minor incommensurate peak, suggesting that 
the volume fraction of the CM phase increases with increasing pressure.

\begin{figure}[htbp]
\begin{center}
\includegraphics[width=49mm]{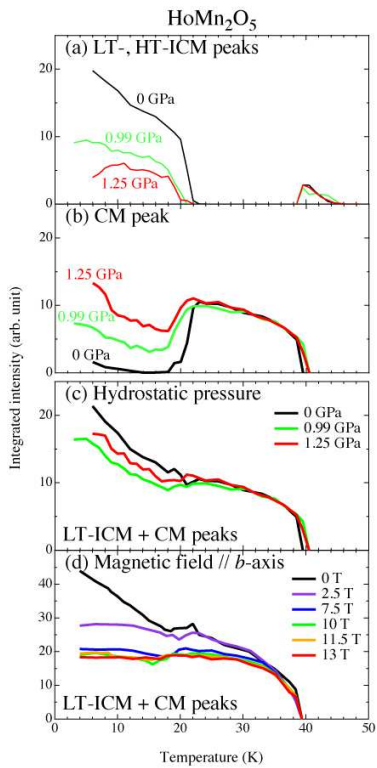}
\end{center}
\caption{
(Color online) 
Temperature dependence of the integrated intensity for the (a) LT-ICM, HT-ICM and 
(b) CM Bragg peaks for HoMn$_{2}$O$_{5}$, taken under $p=0$, 0.99, and 1.25~GPa. 
The sums of the intensities for the LT-ICM and CM peaks as a function of temperature 
under pressure and magnetic field are shown in (c) and (d), respectively. 
The data under magnetic field is taken from Kimura {\it et al}\cite{Kimura2006}.
}
\label{fig3}
\end{figure}

Figures~\ref{fig3} (a) and (b) show the integrated intensities 
of the magnetic peak for the HT-ICM, CM, and LT-ICM phases as a function of temperature,
taken at three different hydrostatic pressures. 
Above $T\sim 22$~K, there is a weak pressure dependence 
of the intensity variation
for both the CM and HT-ICM phases. 
Below $T\sim 22$~K, the behavior of the order parameter for the CM and LT-ICM phases changes significantly upon applying pressure. With increasing pressure, the signal from the LT-ICM phase decreases, and the temperature at which the LT-ICM phase appears slightly decreases. 
In contrast, the intensity coming from the CM phase restores with increasing pressure below 20~K. 
This indicates that the volume fraction 
of the LT-ICM phase reduces and that of the CM phase builds up 
as the applied pressure increases. 
The total amplitude of the magnetic scattering intensity below $\sim 40$~K, which corresponds to 
the sum of the magnetic intensities for the LT-ICM and CM phases,
is plotted as a function of temperature in Fig.~\ref{fig3}(c). The temperature evolution 
of the magnetic intensity at each pressure is almost consistent, where
the intensity rises rapidly below around 22~K. 
This rapid increase might have to do with 
the increase of the magnetic polarization 
of the 4$f$-magnetic moment 
on the Ho$^{3+}$ ion. Fig.~\ref{fig3}(d) shows a similar plot as a function of temperature
obtained under an applied magnetic field,
where the CM phase is also induced by applying a magnetic field\cite{Kimura2006}. 
With increasing magnetic field, the sum of the intensity below $\sim 20$~K rapidly decreases and 
the temperature evolution almost saturates above $H=2.5$~T, while the intensity above 
$\sim 20$~K is almost temperature-independent. This means that 
the sum rule in the total magnetic intensity is broken by applying a magnetic field, 
which sharply contrasts the behavior under hydrostatic pressure. 
These results indicate that the magnetic structure of the CM phase induced by a
magnetic field is different from that induced by hydrostatic pressure, where the 
difference in the structure of Ho$^{3+}$ moments might become dominant. 
\begin{figure}[htbp]
\begin{center}
\includegraphics[width=60mm]{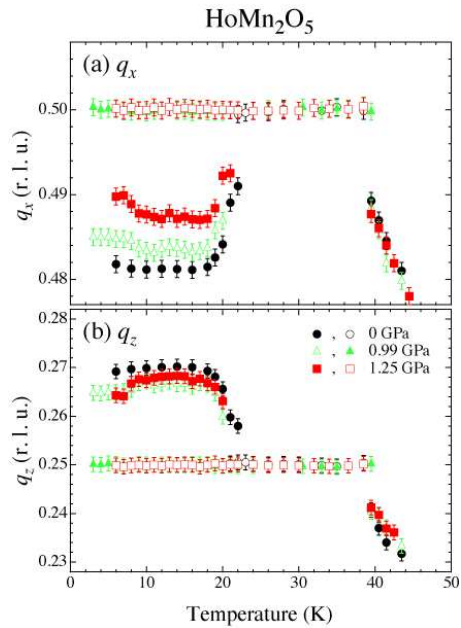}
\end{center}
\caption{
(Color online) 
Temperature dependence of magnetic propagation wave vector ${\bf q}_{\rm M}=(q_{x}\ 0\ q_{z})$ 
of HoMn$_{2}$O$_{5}$ taken at 
$p=0$, 0.99, and 1.25~GPa, denoted by circles, triangles, and squares, respectively. 
Variations of $q_{x}$ and $q_{z}$ components are shown in (a) and (b), respectively. 
}
\label{fig4}
\end{figure}

Temperature dependence of the magnetic propagation wave vector ${\bf q}_{\rm M}=(q_{x}\ 0\ q_{z})$ 
at three different pressure values was measured. 
Figures~\ref{fig4}(a) and (b) show the temperature dependences of 
$q_{x}$ and $q_{z}$, respectively. Above $T\sim 22$~K, the wave vector in both the CM 
and HT-ICM phase is almost independent of the applied pressure, which is consistent with the 
behavior of the order parameters shown in Figs.~\ref{fig3}(a) and (b). 
On the other hand, the wave vector of the LT-ICM phase below 
$T\sim 22$~K strongly depends on the applied pressure. 
The pressure effect becomes enhanced in the behavior of $q_{x}$, which
approaches the commensurate value of 1/2 with increasing pressure. 
This indicates that the instability of the LT-ICM phase increases with 
increasing pressure. 
These results are qualitatively consistent with the results obtained under an applied magnetic field\cite{Kimura2006}. 
Although the two-phase coexistence of the CM and LT-ICM phases
persists up to $p=1.25$~GPa, the critical pressure above which the magnetic phase 
becomes single-phase CM can be estimated by extrapolation from the pressure dependence of 
$q_{x}$. 

\begin{figure}[tbp]
\begin{center}
\includegraphics[width=70mm]{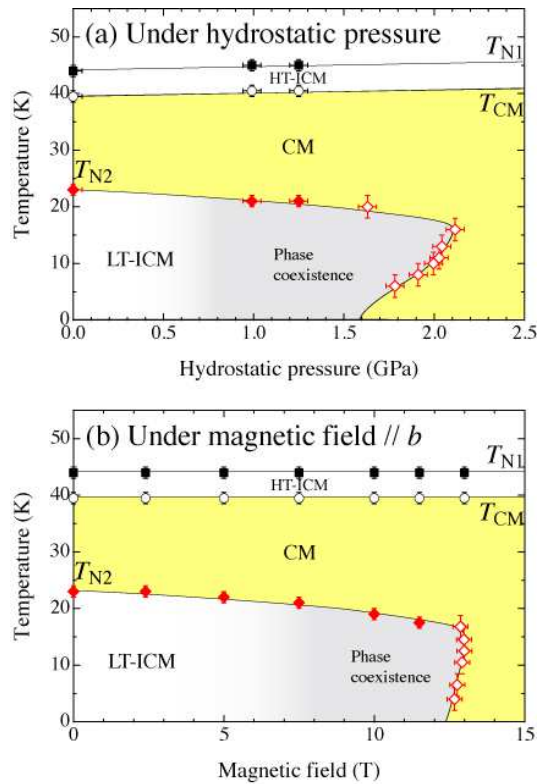}
\end{center}
\caption{
(Color online) 
(a) Pressure -- Temperature magnetic phase diagram for the LT-ICM, CM, and HT-ICM phases 
in HoMn$_{2}$O$_{5}$. (b) Magnetic field -- Temperature phase diagram 
for all the magnetic phases in HoMn$_{2}$O$_{5}$, taken from Kimura {\it et al}.\cite{Kimura2006}. 
Open diamonds were estimated from 
the behavior of $q_{x}$.}
\label{fig5}
\end{figure}

The pressure ($p$) -- temperature ($T$) phase diagram for all the magnetic phases is shown in 
Fig.~\ref{fig5}(a). The magnetic field ($H$) -- temperature ($T$) magnetic phase diagram obtained in 
our 
previous study\cite{Kimura2006} is shown in Fig.~\ref{fig5}(b) for comparison. 
The $p$ -- $T$ phase diagram clearly indicates that the HT-ICM and CM phases at ambient pressure 
become resistant against pressure, and $T_{\rm N1}$ and $T_{\rm CM}$ slightly increase with 
increasing pressure. By contrast, the LT-ICM phase at ambient pressure 
becomes unstable with increasing pressure and finally the system becomes single-phase CM 
above $p\sim 2$~GPa through phase coexistence. Recent study of dielectric measurements 
under hydrostatic pressure in HoMn$_{2}$O$_{5}$ 
revealed that the weak electric polarization in the
X phase grows upon applying pressure\cite{Cruz2007}. 
A comparison of our $p$ -- $T$ magnetic phase diagram 
with the $p$ -- $T$ dielectric phase diagram\cite{Cruz2007}
reveals that there is a one-to-one correspondence between 
the pressure-induced commensurate spin state and the pressure-induced
electric polarization, which is also seen in the
$H$ -- $T$ phase diagram\cite{Higashiyama2005,Kimura2006}. 
This suggests that 
commensurate magnetism leads to bulk electric polarization. 

It is interesting to note that, as seen in Figs.~\ref{fig5}, the $p$ -- $T$ magnetic phase diagram is quite similar 
to the $H$ -- $T$ magnetic phase diagram. 
The magnetic field can directly affect the spins, 
especially the polarization of Ho moments. 
In contrast, pressure can change the interatomic distance or 
bond angles. 
Both the cases change the magnetic interaction and the valance of 
the competing multiple ground states, which can cause magnetic phase transition. 
The present study has demonstrated that the magnetic structure is controlled by not only the magnetic field but also the hydrostatic pressure. However, the reason why the 
phase diagrams obtained under a magnetic field and under pressure are quite similar remains
unclear. The temperature evolution of total magnetic intensity as a function of applied pressure
is different from that as a function of applied magnetic field, as shown in Figs.~\ref{fig3}(c) and (d). 
This suggests that the magnetic structure 
induced by pressure is different from that induced by magnetic field, 
even though both magnetic structures induced 
have the same commensurate periodicity. 
The volume of the unit cell in the LT-ICM phase is larger than that in the CM phase at ambient 
pressure\cite{Blake2005}, showing that the compressed lattice favors the CM phase in the $R$Mn$_{2}$O$_{5}$ 
system. In fact, electric polarization can be induced even in 
TbMn$_{2}$O$_{5}$ and DyMn$_{2}$O$_{5}$\cite{Cruz2007} despite
the difference in rare earth ions, which is indicative of the pressure-induced
CM phase. Crystal as well as magnetic structure analyses under hydrostatic pressure are crucial
for determining what kind of magnetic exchange interaction plays an essential role in
stabilizing the LT-ICM and CM phases. 

The authors acknowledge Dr. K. Kohn for providing a high-quality single crystal of HoMn$_{2}$O$_{5}$. 
We also thank K. Ohoyama and H. Hiraka for their technical assistance in the measurements at AKANE. 
Neutron diffraction measurements 
were performed under the PACS No.~7736 of Japan Atomic Energy Agency. 
This work was supported by Grant-In-Aid for Scientific Research (B) 
(16340096), and Grant-in-Aid for 
Scientific Research on Priority Areas 
``Novel States of Matter Induced by Frustration'' (19052001), 
from the Japanese Ministry of
Education, Culture, Sports, Science and Technology. 
This work was performed under
the Interuniversity Cooperative Research Program, No.
30, of the Institute for Materials Research, Tohoku University. 
\begin{acknowledgments}
\end{acknowledgments}


\end{document}